\documentclass{article}
\usepackage{amsfonts,amssymb, amsmath}
\usepackage[english]{babel}

\def\nn{\nonumber }
\def\bq{ \begin{equation}}
\def\eq{ \end{equation}}
\def\ben{ \begin{eqnarray}}
\def\en{ \end{eqnarray}}

\newtheorem{prop}{Proposition}

\begin{document}


\title{On two-dimensional Hamiltonian systems with  sixth-order integrals of motion}
\author{E.O. Porubov,  A.V. Tsiganov\\
\it\small
St.Petersburg State University, St.Petersburg, Russia\\
\it\small e-mail: st077283@student.spbu.ru, andrey.tsiganov@gmail.com}

\date{}
\maketitle

\begin{abstract}
We obtain  21 two-dimensional natural Hamiltonian systems with sextic invariants, which are polynomial of the sixth order in momenta. Following to Bertrand, Darboux, and Drach
these results of the standard brute force experiments can be applied to construct a new mathematical theory.
 \end{abstract}

\textbf{Keywords:}  integrable systems

\section{Introduction}
\setcounter{equation}{0}
A Hamiltonian system is called natural if Hamiltonian is the sum of positive-definite kinetic energy and potential
\[H=T(q,p)+V(q)\,.\]
The search of natural Hamiltonian systems  admitting integral of motion that is a polynomial of order $N$ in the momenta is a classical problem \cite{darb,darb2}, which is currently intensively studied, see  \cite{mir18,kiy01,koz,km12,sw15,val21} and references within.

There is  few two-dimensional natural Hamiltonian systems with invariants  (constants of motion or integrals of motion), which are polynomials of the sixth order in momenta.
 According  to \cite{hiet87}  the Hamiltonian vector field with Hamiltonian
 \[H=p_1^2+p_2^2+V(q_1,q_2)\,,\]
 admits sextic invariants for a potential  of the Holt-type system
 \[
 V=12q_2^{4/3}+(q_1^2+a)q_2^{-2/3}+bq_1^{-2}\,,
 \]
Toda-type system
\[
V_1=e^\frac{(\sqrt{3}q_1-q_2)}{2}+e^{q_2}+e^{-\sqrt{3}q_1}\,,\qquad
V_2=e^\frac{(\sqrt{3}q_1-q_2)}{2}+e^{q_2}+e^\frac{-\sqrt{3}q_1}{3}\,,
\]
 superintegrable Calogero-type system
 \[
 V=v(2q_2)+v(\sqrt{3}q_1+q_2)+v(-\sqrt{3} q_1+q_2)\,,\qquad v(z)=z^{-2}+\frac{a^2}{2}z^{-2}\,,
 \]
 and a few superintegrable systems associated with the Chebychev theorem on binomial differentials.

 In \cite{ts18} we obtained one more system with  sextic invariant at
\[V=(q_1^{-2/3}+a)q_2^{-2/3}\,,\]
which is a generalization of the Fokas-Lagerstr\"{o}m system having a cubic invariant  \cite{fl80,hiet87}.

On the pseudo-Euclidean space, there are more natural Hamiltonian systems with sextic invariant \cite{ts11}
\[H=p_1p_2+q_1^{a}q_2^{b}\,,\]
where $a$ and $b$ are rational numbers associated with a special set of parameters of the hypergeometric function, for instance
\[V=q_1^2q_2^{-10/7}\,,\qquad V=q_1^{-2/3}q_2^{-5/6}\,,\qquad V=q_1^{-2/3}q_2^{-7/3}\,.\]
Similar to the Holt and Fokas-Lagerstrom potentials we have singular potentials which are well-defined functions only on  part of the plane.

 In  \cite{ts17} we  construct one more Hamiltonian system with  invariant of degree six on the two-dimensional space with  metric depending on local coordinates
\[
H=\dfrac{p_1^2}{2m_1(q_1,q_2)}+\dfrac{p_2^2}{2m_2(q_1,q_2)}+V(q_1,q_2)\,,
\]
 i.e. a natural Hamiltonian system with  position-dependent mass (effective mass). This system has been obtained using divisor arithmetic on an elliptic curve \cite{ts17,ts19a}. Below we present 20 similar natural Hamiltonian systems with effective mass and sextic invariant.

Dynamical systems with position-dependent mass were first introduced in the theory of semiconductor physics, and now these systems can be found in many fields, such as classical Hamiltonian and non-Hamiltonian mechanics,  quantum mechanics, relativistic mechanics, nuclear physics, molecular physics, neutrino mass oscillations, quantum information and so on,  see references in recent papers \cite{bal17,cos20,gr16,pdm21,ts20a}.

Our aim is not to construct integrable systems with a sixth degree invariant or to study the applications of these systems in physics. Similar to Darboux \cite{darb2} and Drach \cite{dr35} we use the direct method of constructing invariants to collect a sufficient number of examples that allow us to build a mathematical theory explaining the existence of such sextic invariants.

\subsection{Two-dimensional geodesic motion with cubic invariant}
Let us consider two-dimensional metric
\[
\mathrm g=\left(
                 \begin{array}{cc}
                   \dfrac{(kq_1+q_2)q_1^m}{q_1-q_2}& 0 \\
                   0 & \dfrac{(kq_2+q_1)q_2^m}{q_2-q_1} \\
                 \end{array}
               \right)\,,\qquad q_1<0<q_2\,,
\]
and the corresponding kinetic energy
\bq\label{gen-ham}
T=\mathrm g_{11} p_1^2+\mathrm g_{22}p_2^2= \dfrac{(kq_1+q_2)q_1^m}{q_1-q_2}\,p_1^2+\dfrac{(kq_2+q_1)q_2^m}{q_2-q_1}\, p_2^2\,,
\eq
derived in \cite{ts17} in the framework of divisor arithmetic on an elliptic curve. Here $q_{1,2}$ are the position variables and  $p_{1,2}$  are the conjugate momenta for canonical Poisson brackets
\[
\{q_1,p_1\}=\{q_2,p_2\}=1\,,\quad \{q_1,q_2\}=\{q_1,p_2\}=\{q_2,p_1\}=\{p_1,p_2\}=0\,.
\]
At $k= \pm 1$ Hamiltonian system with geodesic Hamiltonian $T$ (\ref{gen-ham}) admits linear invariant and, therefore, sextic invariant may be obtained by using the Chebyshev theorem on binomial differentials \cite{ts18}. Here, we do not consider this construction of integrals of motion.

\begin{prop}
At special values of parameters $m$ and $k$
\[\bullet\,\, m=1\,, k=2;\qquad \bullet\,\,m=3\,, k= \frac12,3;\qquad \bullet\,\,m=4\,, k= \pm 3,-\dfrac35,-\dfrac{1}{7},\dfrac{1}{5},\dfrac{1}{2};\]
geodesic Hamiltonian $T$ (\ref{gen-ham}) commutes
\[
 \{T, K\}=0
\]
with the reducible cubic  polynomial
\[K=K_1K_2\,,\qquad K_1=(q_1^2p_1 -q_2^2 p_2)\,,\]
where  $K_2$ is a polynomial of the second order in momenta.
\end{prop}
These special values of parameters are related to  divisor arithmetic on an elliptic curve, which allows us to get separated variables for
the corresponding  Hamilton-Jacobi equation $T=E$ and prove that these Hamiltonian systems are bi-Hamiltonian.

\begin{prop}
At  special values of parameters $m$ and $k$
\[\bullet\,\, m=1\,, k=2;\qquad \bullet\,\,m=3\,, k= \frac12,2;\qquad \bullet\,\,m=4\,, k= -3,\dfrac13,\dfrac15,\dfrac{3}{5},2;\]
geodesic Hamiltonian $T$ (\ref{gen-ham}) commutes
\[
 \{T,L\}=0
\]
 with the reducible quartic  polynomial
\[L=K_1^2L_2\,,\qquad K_1=(q_1^2p_1 -q_2^2 p_2)\,,\]
where $L_2$ is a  polynomial of the second order in momenta.
\end{prop}
Thus, we have superintegrable geodesic motion  at
\[\bullet\,\, m=1\,, k=2;\qquad \bullet\,\,m=3\,, k= \frac12;\qquad \bullet\,\,m=4\,, k= -3,\dfrac15;\]
and in all these cases
\[\{L,K\}= K^2\,.\]

Following the example from  \cite{ts17}, below we consider  integrable Hamiltonian systems with integrals of motion
\bq\label{H}
\begin{array}{rcl}
H_1&=&T+V(q_1,q_2)\,,
\\ \\
H_2&=&\Bigl(K_1^2+U_1(q_1,q_2)\Bigr)\Bigr(K_2+U_2(q_1,q_2)\Bigr)^2\\ \\
&=&
K^2 +  \Bigl(U_1K_2^2+2U_2K_1K\Bigr)+\Bigl(U_2^2K_1^2 +  2U_1U_2K_2\Bigr)+ U_1U_2^2\,.
\end{array}
\eq
where $V(q_1q_2)$ and $U_{1,2}(q_1q_2)$ are the solution of partial differential equations obtained from the standard equation
\bq\label{m-eq}
\{H_1,H_2\}=0\,,
\eq
 which must be identically satisfied for all admissible values of $p_1$ and $p_2$.

If reducible cubic polynomial $K$ admits a few decompositions
\[K=K_1K_2= \hat{K}_1\hat{K}_2=\tilde{K}_1\tilde{K_2}\,,\]
we obtain several integrable systems with integrals of motion (\ref{H}) and
\bq\label{hH}
\begin{array}{rcl}
\hat{H}_1&=&T+\hat{V}(q_1,q_2)\,,\qquad \hat{H}_2=\Bigl(\hat{K}_1^2+\hat{U}_1(q_1,q_2)\Bigr)\Bigr(\hat{K}_2+\hat{U}_2(q_1,q_2)\Bigr)^2\,, \\
\\
\tilde{H}_1&=&T+\tilde{V}(q_1,q_2)\,,\qquad \tilde{H}_2=\Bigl(\tilde{K}_1^2+\tilde{U}_1(q_1,q_2)\Bigr)\Bigr(\tilde{K}_2+\tilde{U}_2(q_1,q_2)\Bigr)^2\,.
\end{array}
\eq
In the next section, we present formal solutions of the partial differential equations (\ref{m-eq}) associated with all parameters $m$ and $k$ in Proposition 1.

\section{Invariants of sixth order in momenta}
\setcounter{equation}{0}
In \cite{ts17} we supposed that $q_{1,2}$ are parabolic coordinates on the plane
\[
q_1 = y - \sqrt{x^2 + y^2}\,,\qquad  q_2 = y + \sqrt{x^2 + y^2}
\]
so that the corresponding momenta are
\[
p_1 = \frac{p_y}{2} +\frac{\sqrt{-q_1q_2}\,p_x}{2q_1},\qquad  p_1 = \frac{p_y}{2} +\frac{\sqrt{-q_1q_2}\,p_x}{2q_2}\,.
\]
In these variables, kinetic energy is equal to
\[
\begin{array}{rcl}
  m=1\,,\qquad T&=&\dfrac{yp_x^2}{2} +\dfrac{ x(k - 1)p_xp_y}{2} + \dfrac{kyp_y^2}{2}\,,\\
  \\
  m=3\,,\qquad T&=& \dfrac{kyx^2p_x^2}{2}+ \dfrac{\left((k - 1)x^2 + 4ky^2\right)xp_xp_y}{2}+\left( \dfrac{(2k-1)x^2}{2} + 2ky^2 \right)yp_y^2\,,\\
  \\
  m=4\,,\qquad T&=&\dfrac{x^2\left((k-1)x^2 + 4ky^2\right) p_x^2}{4} +xy\left((2k-1)x^2 + 4ky^2\right)p_xp_y \\
  \\&+& \left(\dfrac{(k - 1)x^4}{4} + (3k - 1)x^2y^2 + 4ky^4\right)p_y^2\,.
 \end{array}
 \]
 Below we also present potentials in these $x,y$-variables in two dimensional configuration space.

\subsection{Case $m=1$ and $k=2$}
In this case geodesic Hamiltonian
\[
T=\frac{q_1(2q_1 + q_2)p_1^2}{q_1 - q_2} +\dfrac{q_2(q_1 + 2q_2)p_2^2}{q_2-q_1}
\]
commutes with cubic and quartic polynomials
\[
K=\frac{ q_1q_2(q_1^2p_1 -q_2^2p_2)(p_1 - p_2)^2 }{ (q_1 -q_2)^3}\,,\quad L= \frac{(q_1^2p_1 -q_2^2p_2)^2(q_1p_1^2 -q_2p_2^2)}{(q_1 -q_2)^3}\,,
\quad \{K,L\}=K^2\,.
\]
Cubic polynomial  $K$  has the form  $K=K_1K_2=\hat{K_1}\hat{K_2}$, where
\[K_1=(q_1^2p_1 -q_2^2 p_2)\,,\qquad\mbox{and}\qquad  \hat{K}_1=\frac{ q_1q_2(p_1 - p_2)}{ (q_1 -q_2)^3}\,.\]
The corresponding functions in (\ref{H}) and (\ref{hH}) are
\[
\begin{array}{rcl}
V&=&\dfrac{a(4q_1^4 + 14q_1^3q_2 + 19q_1^2q_2^2 + 14q_1q_2^3 + 4q_2^4)}{(q_1 + q_2)^2} +b (q_1 + q_2) +c\left(\dfrac{1}{q_1}+\dfrac{1}{q_2}\right)\,,
\\ \\
U_1&=&2a(q_1^2 + q_1q_2 + q_2^2)(q_1 + q_2)(q_1 - q_2)^2 +\dfrac{b(q_1 - q_2)^2(q_1 + q_2)^2}{2}\,,
\\ \\
U_2&=&\dfrac{2aq_1q_2}{(q_1 + q_2)(q_1 - q_2)} -\dfrac{c}{q_1q_2(q_1 - q_2)}\,,
\end{array}
\]
and
\[
\begin{array}{rcl}
\hat{V}&=&\dfrac{a(q_1 ^4 - 4 q_1 ^3 q_2  - 14 q_1 ^2 q_2 ^2 - 4 q_1  q_2 ^3 + q_2 ^4)}{(q_1  + q_2 )^2} +\dfrac{b (q_1 ^2 + 3 q_1  q_2  + q_2 ^2)}{(q_1  + q_2 )^{3/2}} +c (q_1  + q_2 )\,,
\\ \\
\hat{U_1}&=&\dfrac{8a q_2 ^2 q_1 ^2}{(q_1^2  - q_2^2 )^4} - \dfrac{2b q_1  q_2}{\sqrt{q_1  + q_2} (q_1  - q_2 )^4}\,,
\\ \\
\hat{U}_2&=&\dfrac{2a (q_1 ^4- q_2^4) }{q_1  + q_2} +\dfrac{b (q_1  - q_2 )^2}{2\sqrt{q_1  + q_2}} + c (q_1  - q_2 )^2\,.
\end{array}
\]
In $x,y$-variables potentials read as
\ben\label{V1}
V&=&-\frac{a(x^4 - 8x^2y^2 - 64y^4)}{4y^2} + 2by - \frac{2cy}{x^2}\,,\\
\nn\\
\label{V2}
\hat{V}&=&-\frac{a(x^4 - 8x^2y^2 - 4y^4)}{y^2} - \frac{\sqrt{2}b(x^2 - 4y^2)}{4y^{3/2}} + 2cy\,.
\en
Using the method of undetermined coefficients we can prove that these Hamiltonian systems do not have quartic invariants. Indeed,
substituting  quartic polynomial
\[H_3=L+\sum_{k=0}^4 \sum_{i=0}^k f_{ik}(q_1,q_2)p_1^{k-i}p_2^i\]
into the equations
\[
\{T+V,H_3\}=0\qquad\mbox\qquad \{T+\hat{V},H_3\}=0
\]
we obtain two inconsistent systems of partial differential equations on functions $f_{ik}(q_1,q_2)$. It means that the corresponding Hamiltonian vector fields admit quartic invariants only at $V=0$ and $\hat{V}=0$.

 \subsection{Case $m=3$ and $k=3$}
 Geodesic Hamiltonian
 \[
  T=\dfrac{q_1 ^3 (3 q_1  + q_2 ) p_1 ^2}{q_1  - q_2 }+\dfrac{ q_2 ^3 (q_1  + 3 q_2 ) p_2^2}{q_2  - q_1 }\,,\\ \\
 \]
 commutes with cubic polynomial
 \[
 K=\frac{q_1^{3/2}q_2^{3/2}}{(q_1-q_2)^2}(q_1^2p_1 -q_2^2 p_2)(q_1^{3/2}p_1 -q_2^{3/2} p_2)(q_1^{3/2}p_1 +q_2^{3/2} p_2)\,.
 \]
 So, we can take
 \[
 K_1=q_1^{3/2}q_2^{3/2}(q_1^2p_1 -q_2^2 p_2)\,,\qquad \hat{K}_1=q_1^{3/2}q_2^{3/2}(q_1^{3/2}p_1 - q_2^{3/2} p_2)\,.
 \]
 and
 \[
 \tilde{K}_1=q_1^{3/2}q_2^{3/2}(q_1^{3/2}p_1 + q_2^{3/2} p_2)\,.
 \]
  Solution of the  equation (\ref{m-eq}) for $V$ and $U_{1,2}$ has the following form
 \[
 \begin{array}{rcl}
 V&=&\dfrac{a(q_1  + q_2 ) (q_1 ^2 + 3 q_1  q_2  + q_2 ^2)}{q_1 ^3 q_2 ^3} +{ b (q_1  + q_2 )} +c\left(\dfrac{1}{q_1}+\dfrac{1}{q_2} \right )\,,
 \\ \\
 U_1&=&\dfrac{a (q_1  - q_2 )^2 (q_1  + q_2 )^2}{4}+\dfrac{b q_1 ^3 q_2 ^3 (q_1  - q_2 )^2}{3} \,,
 \\ \\
 U_2&=&\dfrac{a(q_1 ^2 + q_1  q_2  + q_2 ^2) }{q_1^3q_2 ^3 (q_1  - q_2 )} +\dfrac{ b}{3 (q_1  - q_2 )} +\dfrac{c}{q_1  q_2  (q_1  - q_2 )}\,.
\end{array}
 \]
 Other decompositions give rise to potentials
 \[
  \begin{array}{c}
 \hat{V}=\frac{a(13 q_1^2 + 4\sqrt{ q_1^3 q_2}+ 46 q_1 q_2 + 4 \sqrt{q_1q_2^3} + 13 q_2^2)\left(\sqrt{q_1}+ \sqrt{q_2}\right)^2}{q_1^3 q_2^3} +\frac{ bq_1 q_2\left(\sqrt{q_1}+ \sqrt{q_2}\right)^2}{(3 q_1 - 2 \sqrt{q_1q_2}+ 3 q_2)^2} +\frac{c\left(\sqrt{q_1}+ \sqrt{q_2}\right)^2}{q_1 q_2}\,,\\
 \\
  \tilde{V}=\frac{a(13 q_1^2 - 4\sqrt{ q_1^3 q_2}+ 46 q_1 q_2 - 4 \sqrt{q_1q_2^3} + 13 q_2^2)\left(\sqrt{q_1}- \sqrt{q_2}\right)^2}{q_1^3 q_2^3} +\frac{ bq_1 q_2\left(\sqrt{q_1}- \sqrt{q_2}\right)^2}{(3 q_1+ 2 \sqrt{q_1q_2}+ 3 q_2)^2} +\frac{c\left(\sqrt{q_1}- \sqrt{q_2}\right)^2}{q_1 q_2}\,,
   \end{array}
 \]
 and functions
 \[
 \begin{array}{l}
 \hat{U}_1=-8a(\sqrt{q_1} - \sqrt{q_2})^2(\sqrt{q_1} +\sqrt{q_2})^4 +\dfrac{8bq_1^4q_2^4(\sqrt{q_1} - \sqrt{q_2})^2}{3(3q_1 - 2\sqrt{q_1q_2} + 3q_2)^2}\,,\\
 \\
 \tilde{U}_1=-8a(\sqrt{q_1} - \sqrt{q_2})^4(\sqrt{q_1} +\sqrt{q_2})^2 -\dfrac{8bq_1^4q_2^4(\sqrt{q_1}+ \sqrt{q_2})^2}{3(3q_1 + 2\sqrt{q_1q_2} + 3q_2)^2}\,,\\
 \\
 \hat{U}_2=-\frac{a(5 q_1 + 2 \sqrt{q_1q_2} + 5 q_2) (q_1 - 6\sqrt{q_1q_2} + q_2)}{(\sqrt{q_1}- \sqrt{q_2}) q_1^3 q_2^3} - \frac{b q_1 q_2}{3 (\sqrt{q_1}- \sqrt{q_2})
 (3 q_1 - 2\sqrt{q_1q_2}+ 3 q_2)^2} +\frac{c}{q_1 q_2 (\sqrt{q_1}- \sqrt{q_2})}\,,\\
 \\
 \tilde{U}_2=-\frac{a(5 q_1 - 2 \sqrt{q_1q_2} + 5 q_2) (q_1+ 6\sqrt{q_1q_2} + q_2)}{(\sqrt{q_1}+ \sqrt{q_2}) q_1^3 q_2^3} + \frac{b q_1 q_2}{3 (\sqrt{q_1}+ \sqrt{q_2})
 (3 q_1+ 2\sqrt{q_1q_2}+ 3 q_2)^2} -\frac{c}{q_1 q_2 (\sqrt{q_1}+ \sqrt{q_2})}\,,
 \end{array}
 \]
In $x,y$-variables potentials have the form
\ben\label{V3}
V&=&\frac{2ay(x^2 - 4y^2)}{x^6} + 2by - \frac{2cy}{x^2}\,, \\
\nn\\
\label{V4}
\hat{V}&=&\mathrm i\left(-\frac{8 a(\mathrm i y - 5x - 8y)(\mathrm i y - 5x + 8y)(\mathrm i y - x)}{5x^6} - \dfrac{b(\mathrm i y - x)x^2}{2(3\mathrm i y + x)^2} - \frac{2c(\mathrm i y - x)}{x^2}\right)\,,\\
\nn\\
\label{V5}
\tilde{V}&=&\mathrm i\left(-\frac{8 a(\mathrm i y + 5x - 8y)(\mathrm i y + 5x + 8y)(\mathrm i y + x)}{5x^6} - \dfrac{b(\mathrm i y + x)x^2}{2(3\mathrm i y - x)^2} - \frac{2c(\mathrm i y + x)}{x^2}\right)\,,
\en
where $\mathrm i=\sqrt{-1}$. In  two last cases canonical transformation of  coordinates
\[y\to\mathrm i y \qquad\mbox{and}\qquad p_y\to -\mathrm i p_y,\]
changes kinetic energy  $T$ and potentials $\hat{V}$, $\tilde{V}$ so that the corresponding Hamiltonians
\[
\hat{H}_1=\mathrm i (T+\hat{ V})\,,\qquad \mbox{and}\qquad\tilde{H}_1=\mathrm i (T+\tilde{ V})
\]
become real functions.
  \subsection{Case $m=3$ and $k=1/2$}
  Quadratic polynomial
  \[
  T=\frac{q_1^3\left(\frac{q_1}{2} + q_2\right)p_1^2}{q_1 -q_2}+\frac{q_2^3\left(q_1 + \frac{q_2}{2}\right)p_2^2}{q_2-q_1}
  \]
  commutes with cubic and quartic polynomials
  \[
  K=\frac{q_1^2q_2^2(q_1^2p_1 -q_2^2p_2)^2(p_1 - p_2)}{(q_1 - q_2)^3}\,,\qquad
  L=\frac{(q_1^2p_1 -q_2^2p_2)^3\Bigl(q_1 ^2 (q_1  + 3 q_2 ) p_1 + q_2 ^2 (3 q_1  + q_2 )p_2\Bigr)}{4(q_1 - q_2)^3}
  \]
  so that $\{L,K\}=K^2$.
 Because cubic polynomial $K$ admits two decompositions  \[K=K_1K_2= \hat{K}_1\hat{K}_2\,,\] where
  \[  K_1= q_1^2q_2^2(q_1^2p_1 -q_2^2 p_2)\,,\qquad\mbox{and}\qquad  \hat{K}_1=\frac{p_1 - p_2}{(q_1 -q_2)^3}\,,\]
  we can construct two different integrable systems with sextic invariants so that
  \[
   \begin{array}{rcl}
   V&=&\dfrac{a(q_1 ^4 - 4 q_1 ^3 q_2  - 14 q_1 ^2 q_2 ^2 - 4 q_1  q_2 ^3 + q_2 ^4)}{q_1 ^2 q_2 ^2 (q_1  + q_2 )^2} +\dfrac{b(q_1 ^2 + 3 q_1  q_2  + q_2 ^2)}{\sqrt{q_1q_2} (q_1  + q_2 )^{3/2}}
    + c\left(\dfrac{1}{q_1}+\dfrac{1}{q_2}\right)\,,
   \\ \\
   U_1&=&-\dfrac{16 a q_1 ^3 q_2 ^3 (q_1  - q_2 )^2}{q_1  + q_2} +\dfrac{ 4 b q_1 ^{7/2} q_2 ^{7/2} (q_1  - q_2 )^2}{\sqrt{q_1  + q_2}}\,,
   \\ \\
   U_2&=&-\dfrac{4a (q_1 ^2 + q_2 ^2)}{q_1 ^3 q_2 ^3 (q_1^2- q_2^2 )} -\dfrac{ b}{q_1^{3/2} q_2^{3/2} \sqrt{q_1+q_2}(q_1^2 - q_2^2)} -\dfrac{2c}{q_1 ^2 q_2 ^2 (q_1  - q_2 )}\,,
   \end{array}
   \]
   and
   \[
   \begin{array}{rcl}
   \hat{V}&=&\dfrac{a(4 q_1 ^4 + 14 q_1 ^3 q_2  + 19 q_1 ^2 q_2 ^2 + 14 q_1  q_2 ^3 + 4 q_2 ^4)}{q_1 ^2 q_2 ^2 (q_1  + q_2 )^2} + b(q_1 + q_2)+ c\left(\dfrac{1}{q_1}+\dfrac{1}{q_2}\right)\,,
   \\ \\
   \hat{U}_1&=&-\dfrac{4a (q_1  + q_2 ) (q_1 ^2 + q_1  q_2  + q_2 ^2)}{q_1^5q_2 ^5 (q_1  - q_2 )^4} -\dfrac{c (q_1  + q_2 )^2}{q_1 ^4 q_2 ^4 (q_1  - q_2 )^4}\,,
   \\ \\
   \hat{U}_2&=&-\dfrac{4a q_1q_2 (q_1  - q_2 )^2}{q_1  + q_2}+2bq_1^2q_2^2(q_1  - q_2 )^2 \,.
   \end{array}
  \]
 In this case
 \ben\label{V6}
 V&=&-\frac{a(x^4 - 8x^2y^2 - 4y^4)}{x^4y^2}+\frac{\mathrm i b\sqrt{2}(x^2 - 4y^2)}{4xy^{3/2}}-\frac{2cy}{x^2}\,, \\
 \nn\\
 \label{V7}
  \hat{V}&=&-\frac{a(x^4 - 8x^2y^2 - 64y^4)}{4x^4y^2} + 2by - \frac{2cy}{x^2}\,.
  \en
 After changing parameter $b\to-\mathrm i b$  we obtain real potential $V$ (\ref{V6}).
 \subsection{Case $m=4$ and $k=-3$}
 In this case geodesic Hamiltonian
 \[
 T= \frac{q_1 ^4 (q_2-3 q_1 ) p_1 ^2}{q_1  - q_2}+ \frac{ q_2 ^4 (q_1  - 3 q_2 ) p_2 ^2}{q_2-q_1}
 \]
 commutes with  the reducible cubic polynomial
 \[K=K_1K_2= \hat{K}_1\hat{K}_2=\frac{q_1q_2(q_1^2p_1 -q_2^2p_2) (q_1^2p_1 + q_2^2p_2)^2}{q_1 - q_2}\,,\]
  where
\[
K_1=q_1^2p_1 -q_2^2p_2 \qquad\mbox{and}\qquad \hat{K}_1=\frac{q_1q_2(q_1^2p_1 + q_2^2p_2)}{q_1 - q_2}\,.\]
The corresponding quartic invariant is equal to
\[
L=\frac{(q_1^2p_1 -q_2^2p_2)^2\Bigl(  q_1 ^4 (3 q_1 ^2 - 2 q_1  q_2  + q_2 ^2) p_1^2 + 2 q_1 ^2 q_2 ^2 (q_1 ^2 + q_2 ^2) p_1  p_2  + q_2 ^4 (q_1 ^2 - 2 q_1  q_2  + 3 q_2 ^2) p_2^2\Bigr)}{4(q_1  - q_2 )^2}\,,
\]
so that $\{L,K\}=K^2$. The corresponding two integrable systems with sextic invariants  (\ref{H}-\ref{hH}) are determined by functions
 \[
 \begin{array}{rcl}
 V&=&\dfrac{a(19 q_1 ^4 - 68 q_1 ^3 q_2  + 82 q_1 ^2 q_2 ^2 - 68 q_1  q_2 ^3 + 19 q_2 ^4)}{q_1 ^4 q_2 ^4} +\dfrac{b q_1 ^2 q_2 ^2}{(q_1  + q_2 )^2}
  +\dfrac{c (q_1  - q_2 )^2}{q_1 ^2 q_2 ^2}\,,
 \\ \\
U_1&=&-\dfrac{8a (3 q_1 ^2 - 2 q_1  q_2  + 3 q_2 ^2) (q_1  - q_2 )^2}{q_1 ^4 q_2 ^4} -\dfrac{c (q_1  - q_2 )^2}{q_1 ^2 q_2 ^2}\,,
 \\ \\
U_2&=&-\dfrac{4a (q_1  + q_2 )^2 (q_1  - q_2 )}{q_1 ^3 q_2 ^3} -\dfrac{b q_1 ^3 q_2 ^3}{(q_1  + q_2 )^2 (q_1  - q_2 )}\,,
 \end{array}
 \]
 and
   \[
   \begin{array}{rcl}
   \hat{V}&=&\dfrac{a(q_1 ^4 - 2 q_1 ^3 q_2  - 2 q_1 ^2 q_2 ^2 - 2 q_1  q_2 ^3 + q_2 ^4)}{q_1 ^4 q_2 ^4}
   +\dfrac{b(q_1 ^2 - 6 q_1  q_2  + q_2 ^2)}{4 q_1 q_2  (q_1  - q_2 )} +\dfrac{c (q_1  - q_2 )^2}{q_1 ^2 q_2 ^2}\,,
   \\ \\
   \hat{U}_1&=&-\dfrac{a(q_1  + q_2 )^2}{q_1 ^2 q_2 ^2} -\dfrac{ b q_1  q_2}{q_1  - q_2}\,,
   \\ \\
   \hat{U}_2&=&\dfrac{a(q_1  + q_2 ) (q_1  - q_2 ) (q_1 ^2 + q_2 ^2)}{q_1 ^4 q_2 ^4} +\dfrac{b(q_1  + q_2 )}{4 q_1  q_2} +{  c}\left(\dfrac{1}{q_1 ^2}-\dfrac{1}{q_2^2}\right)\,.
   \end{array}
   \]
  In $x,y$-variables potentials have the form
  \ben\label{V8}
 V&=&\frac{16a(16x^4 + 36x^2y^2 + 19y^4)}{x^8} +\frac{b x^4}{4y^2} + \frac{4c(x^2 + y^2)}{x^4}\,,\\
\nn\\
\label{V9}
   \hat{V}&=&\frac{4a(x^4 + 6x^2y^2 + 4y^4)}{x^8} +\frac{b(2x^2 + y^2)}{2x^2\sqrt{x^2 + y^2}} + \frac{4c(x^2 + y^2)}{x^4}\,.
\en
We can  prove that the  Hamiltonians $H_1$ (\ref{H})  and $\hat{H}_1$ (\ref{hH})  do not commute with the following polynomial of fourth order in momenta
\[H_3=L+\sum_{k=0}^4 \sum_{i=0}^k f_{ik}(q_1,q_2)p_1^{k-i}p_2^i\,,\]
for any functions $f_{ik}(q_1,q_2)$.  It means that the corresponding Hamiltonian vector fields admit quartic invariants only for geodesic motion at $V=0$ and $\hat{V}=0$.
 \subsection{Case $m=4$ and $k=3$}
 In this case, we have
 \[
T=\frac{q_1^4(3q_1 + q_2)p_1^2}{q_1 - q_2}+\frac{q_2^4(q_1 + 3q_2)p_2^2}{q_2-q_1}
 \]
 and
 \[
 K=K_1K_2=\hat{K}_1\hat{K}_2= \frac{q_1^2q_2^2}{(q_1-q_2)^2}(q_1^2p_1 -q_2^2p_2)^2(q_1^2p_1 +q_2^2 p_2)\,,
 \]
 where
 \[ K_1=q_1^2q_2^2(q_1^2p_1 -q_2^2p_2)\qquad\mbox{and}\qquad\hat{K_1}=(q_1^2p_1 +q_2^2 p_2)\,.\]
  The first decomposition yields  the following solutions of (\ref{m-eq})
 \[
 \begin{array}{rcl}
 V&=&\dfrac{a(q_1 ^4 + 4 q_1 ^3 q_2  + 4 q_1 ^2 q_2 ^2 + 4 q_1  q_2 ^3 + q_2 ^4)}{q_1 ^4 q_2 ^4} +\dfrac{b (q_1 ^2 + 4 q_1  q_2  + q_2 ^2)}{q_1 ^2 q_2^2}+c\left(\dfrac{1}{q_1}+\dfrac{1}{q_2}\right)\,,
 \\ \\
 U_1&=&-\dfrac{a(3 q_1 ^2 + 2 q_1  q_2  + 3 q_2 ^2) (q_1  - q_2 )^2}{4} - b q_2 ^2 q_1 ^2 (q_1  - q_2 )^2\,,
 \\ \\
 U_2&=&\dfrac{a(q_1  + q_2 ) (q_1 ^2 + q_2 ^2)}{q_1 ^4 q_2 ^4 (q_1  - q_2 )} +\dfrac{b (q_1  + q_2 )}{q_1 ^2 q_2 ^2 (q_1  - q_2 )} +\dfrac{c}{q_1  q_2  (q_1  - q_2 )}\,.
 \end{array}
 \]
The second decomposition leads to the solutions
\[
 \begin{array}{rcl}
 \hat{V}&=&\dfrac{a(q_1^2 + 6 q_1 q_2 + q_2^2) (q_1 - q_2)^2}{q_1^4 q_2^4} +\dfrac{b q_1^2 q_2^2}{(q_1 + q_2)^2} +\dfrac{c (q_1 - q_2)^2}{q_1^2 q_2^2}\,,
 \\ \\
 \hat{U}_1&=&-\dfrac{a(q_1 - q_2)^4}{q_1^4 q_2^4} + \dfrac{2b q_1^2 q_2^2}{(q_1 + q_2)^2}\,,
 \\ \\
 \hat{U}_2&=&-\dfrac{2a (3 q_1^2 + 2 q_1 q_2 + 3 q_2^2)}{q_1^2 q_2^2} + 2c\,.
 \end{array}
 \]
In $x,y$-variables potentials have the form
  \ben\label{V10}
 V&=&-\frac{2a(x^4 - 8y^4)}{x^8} -\frac{ 2b(x^2 - 2y^2)}{x^4} - \frac{2cy}{x^2 }\,,\\
\nn\\
\label{V11}
   \hat{V}&=&-\frac{16a(x^4 - y^4)}{x^8} + \frac{bx^4}{4y^2} +\frac{4c (x^2 + y^2)}{x^4}\,.
\en
 \subsection{Case $m=4$ and $k=-3/5$}
 Geodesic Hamiltonian
 \[
 T=\frac{q_1^4\left(q_2-\frac{3q_1}{5}\right)p_1^2}{q_1-q_2}+ \frac{q_2^4\left(q_1 -\frac{3q_2}{5}\right)p_2^2}{q_2-q_1}
 \]
 commutes with reducible polynomial
 \[K=\frac{(q_1^2p_1 -q_2^2p_2)(q_1^2p_1 +q_2^2 p_2)\bigl(q_1^2(5q_1^2 - 10q_1q_2 + q_2^2)p_1
 + q_2^2(q_1^2 - 10q_1q_2 + 5q_2^2)p_2\bigr)}{q_1q_2(q_1 -q_2)}\,,\]
 from which we can extract three different  linear polynomials in momenta
 \[
 K_1=(q_1^2p_1 -q_2^2p_2)\,,\qquad \hat{K}_1=\frac{(q_1^2p_1 +q_2^2 p_2)}{q_1q_2(q_1 - q_2)}\]
 and
 \[\tilde{K}_1=q_1^2(5q_1^2 - 10q_1q_2 + q_2^2)p_1 + q_2^2(q_1^2 - 10q_1q_2 + 5q_2^2)p_2\,.
 \]
The corresponding potentials in (\ref{H}-\ref{hH}) are equal to
\[
\begin{array}{rcl}
V&=&\dfrac{a(19 q_1 ^2 - 26 q_1  q_2  + 19 q_2 ^2)}{q_1 ^2 q_2 ^2} +\dfrac{b q_1 ^2 q_2 ^2}{(q_1  + q_2 )^2}\,,
\\ \\
U_1&=&-\dfrac{40 a (q_1  - q_2 )^2}{q_1 ^2 q_2 ^2}\,,
\\ \\
U_2&=&\dfrac{5a (5 q_1 ^2 - 6 q_1  q_2  + 5 q_2 ^2) (q_1  + q_2 )^2}{2q_1 ^3 q_2 ^3 (q_1  - q_2 ) } - \dfrac{10bq_1 q_2  (q_1  - q_2 )}{(q_1  + q_2 )^2}
\end{array}
\]
and
\[
\begin{array}{l}
\hat{V}=\dfrac{a(31 q_1 ^4 - 68 q_1 ^3 q_2  + 58 q_1 ^2 q_2 ^2 - 68 q_1  q_2 ^3 + 31 q_2 ^4)}{q_1q_2(q_1  - q_2 )^3}
 -\dfrac{b\sqrt{q_1q_2} (7 q_1 ^2 - 18 q_1  q_2  + 7 q_2 ^2)}{(q_1  - q_2 )^{5/2}} +\dfrac{c q_1  q_2}{q_1  - q_2 }\,,
\\ \\
\hat{U}_1=-\dfrac{40 a (q_1  + q_2 )^2}{q_1 ^3 q_2 ^3 (q_1  - q_2 )^3}+\dfrac{40 b}{q_1 ^{3/2}q_2^{3/2} (q_1  - q_2 )^{5/2} } \,,
\\ \\
\hat{U}_2=\dfrac{5a (q_1  + q_2 ) (3 q_1  - q_2 )^2 (q_1  - 3 q_2 )^2}{q_1q_2(q_1  - q_2 )^2} + \dfrac{5b \sqrt{q_1q_2}) (3 q_1  - q_2 ) (q_1  - 3 q_2 ) (q_1  + q_2 )}{(q_1  - q_2 )^{3/2}}\\
\\
- 5 c q_1q_2 (q_1  + q_2 ) \,.
\end{array}
\]
Third decomposition $K=\tilde{K}_1\tilde{K}_2$ yields the third integrable system $\tilde{H}_1=T+\tilde{V}$ with potential
\[
\tilde{V}=\frac{a (5 q_1 ^2 - 6 q_1  q_2  + 5 q_2 ^2)}{q_1q_2  (q_1  - q_2 )}\,,\qquad \tilde{U}_1 =0\,,\qquad \tilde{U}_2=\frac{5a(q_1  + q_2 )}{q_1  q_2 }\,.
\]
Thus, in this case, we have three integrable systems with sextic invariants and potentials
\ben \label{V12}
V&=&\frac{4a(16x^2 + 19y^2)}{x^4} +\frac{bx^4}{4y^2}\,,
\\ \nn\\
\label{V13}
\hat{V}&=&\frac{2a(16x^4 + 48x^2y^2 + 31y^4)}{x^2(x^2 + y^2)^{3/2}} +\frac{\sqrt{2}b x(8x^2 + 7y^2)}{2(x^2 + y^2)^{5/4}} -\frac{c x^2}{2\sqrt{x^2 + y^2}}\,,
\\ \nn\\
\label{V14}
\tilde{V}&=&\frac{2a(4x^2 + 5y^2)}{x^2\sqrt{x^2 + y^2}}\,.
\en
 \subsection{Case $m=4$ and $k=-1/7$}
 As above geodesic Hamiltonian
 \[
 T=\frac{q_1^4\left(q_2-\frac{q_1}{7}\right)p_1^2}{q_1-q_2}+ \frac{q_2^4\left(q_1 -\frac{q_2}{7}\right))p_2^2}{q_2-q_1}
 \]
 commutes with reducible cubic  polynomial
  \[\begin{array}{rcl}
  K&=&\dfrac{(q_1^2p_1 -q_2^2p_2)\Bigl(q_1^2(q_1 - 3q_2)p_1 + q_2^2(3q_1 -q_2)p_2\Bigr)}{q_1q_2(q_1 - q_2)^3}\\
  \\
  &\times&\Bigl(q_1^2(3q_1^3 - 27q_1^2q_2 + 33q_1q_2^2 - q_2^3)p_1 + q_2^2(q_1^3 - 33q_1^2q_2 + 27q_1q_2^2 - 3q_2^3)p_2\Bigr)
  \end{array}
  \]
  from which we can extract three different  linear polynomials
 \[
 K_1=q_1^2p_1 -q_2^2p_2\,,\qquad \hat{K}_1=q_1^2(q_1 - 3q_2)p_1 + q_2^2(3q_1 -q_2)p_2\]
 and
 \[\tilde{K}_1=q_1^2(3q_1^3 - 27q_1^2q_2 + 33q_1q_2^2 - q_2^3)p_1 + q_2^2(q_1^3 - 33q_1^2q_2 + 27q_1q_2^2 - 3q_2^3)p_2\,.
 \]
The corresponding potentials in (\ref{H}-\ref{hH}) are equal to
\[
\begin{array}{rcl}
V&=&\dfrac{a(11 q_1 ^2 - 42 q_1  q_2  + 11 q_2 ^2)}{q_1 ^2 q_2 ^2} +\dfrac{b q_1 ^2 q_2 ^2}{(q_1  + q_2 )^2}\,,\quad U_1=\dfrac{56 a (q_1  - q_2 )^2}{q_1 ^2 q_2 ^2}\,,
\\ \\
U_2&=&\dfrac{14a (3 q_1  - q_2 ) (q_1  - 3 q_2 ) (q_1  + q_2 )^2}{q_1^3q_2^3(q_1  - q_2 )} -\dfrac{28b q_1  q_2  (q_1 ^2 - 6 q_1  q_2  + q_2 ^2)}{(q_1  + q_2 )^2 (q_1  - q_2 )}\,,
\end{array}
\]
  and
 \[
 \begin{array}{rcl}
 \hat{V}&=&\dfrac{a(11 q_1 ^4 - 84 q_1 ^3 q_2  + 66 q_1 ^2 q_2 ^2 - 84 q_1  q_2 ^3 + 11 q_2 ^4)}{q_1q_2  (q_1 ^2 - 6 q_1  q_2  + q_2 ^2)^{3/2}}
  - \dfrac{2b q_1  q_2}{\sqrt{q_1 ^2 - 6 q_1  q_2  + q_2 ^2}}\,,
 \\ \\
 \hat{U}_1&=&-\dfrac{112 a (q_1  - q_2 )^2 (q_1  + q_2 )^2}{q_1q_2\sqrt{q_1 ^2 - 6 q_1  q_2  + q_2 ^2}}\,,
 \\ \\
 \hat{U}_2&=&\dfrac{14a (q_1  + q_2 ) (q_1 ^2 - 14 q_1  q_2  + q_2 ^2)^2}{(q_1 ^2 - 6 q_1  q_2  + q_2 ^2)^{3/2} (q_1  - q_2 ) q_1 ^2 q_2 ^2}
  +\dfrac{ 28b (q_1  + q_2 )}{\sqrt{q_1 ^2 - 6 q_1  q_2  + q_2 ^2} (q_1  - q_2 )}\,.
  \end{array}
 \]
Third decomposition $K=\tilde{K}_1\tilde{K}_2$ gives rise to the third integrable system so that
\[
\tilde{V}=\frac{a(3 q_1  - q_2 ) (q_1  - 3 q_2 )}{ q_1  q_2 \sqrt{q_1 ^2 - 6 q_1  q_2  + q_2 ^2}}\,,\qquad \tilde{U}_1 =0\,,\qquad \tilde{U}_2=\dfrac{14a (q_1  + q_2 )}{
 q_1 ^2 q_2 ^2 (q_1  - q_2 )\sqrt{q_1 ^2 - 6 q_1  q_2  + q_2 ^2}}\,.
\]
In $x,y$-variables these potentials have the form
\ben\label{V15}
V&=&\frac{4a(16x^2 + 11y^2)}{x^4} +\frac{bx^4}{4y^2},,\\
\nn\\
\label{V16}
\hat{V}&=&-\frac{2a(16x^4 + 32x^2y^2 + 11y^4)}{x^2(2x^2 + y^2)^{3/2}} +\frac{b x^2}{\sqrt{2x^2 + y^2}}\,,\\
\nn\\
\label{V17}
\tilde{V}&=&-\frac{2a(4x^2 + 3y^2)}{x^2\sqrt{2x^2 + y^2}}\,.
\en

\subsection{Case $m=4$ and $k=1/5$}
In this case geodesic Hamiltonian
\[
T=\frac{q_1^4\left(\frac{q_1}{5} + q_2\right)p_1^2}{q_1 -q_2} +\frac{q_2^4\left(q_1 +\frac{q_2}{5}\right)p_2^2}{q_2-q_1}
\]
commutes with  the cubic and quartic polynomials
\[K=\frac{q_1q_2(q_1^2p_1 -q_2^2p_2)\Bigl(q_1^2(q_1 + 3q_2)p_1 - q_2^2(3q_1 + q_2)p_2\Bigr)^2}{(q_1 -q_2)^3}\,,\]
and
\[
L=\scriptstyle \frac{ (q_1^2p_1 -q_2^2p_2)^2\bigl(
q_1 ^4 (q_1 ^3 + 9 q_1 ^2 q_2  + 21 q_1  q_2 ^2 + q_2 ^3) p_1^2 - 2 q_1 ^2 q_2 ^2 (q_1  - q_2 ) (q_1 ^2 + 4 q_1  q_2  + q_2 ^2) p_1p_2 - q_2 ^4 (q_1 ^3 + 21 q_1 ^2 q_2  + 9 q_1  q_2 ^2 + q_2 ^3) p_2^2
\bigr)}{4(q_1  - q_2 )^3}\,.
\]
so that $\{L,K\}=K^2$.

This cubic polynomial can be decomposed as $K=K_1K_2$ or $K=\hat{K}_1\hat{K}_2$, where
\[
K_1=(q_1^2p_1 -q_2^2p_2)\qquad \mbox{and}\qquad
\hat{K}_1=q_1^2(q_1 + 3q_2)p_1 - q_2^2(3q_1 + q_2)p_2\,.
\]
Integrals of motion (\ref{H}-\ref{hH}) involve the following functions
\[
\begin{array}{l}
V=\dfrac{a(11 q_1 ^4 + 156 q_1 ^3 q_2  + 546 q_1 ^2 q_2 ^2 + 156 q_1  q_2 ^3 + 11 q_2 ^4)}{q_1 ^4 q_2 ^4} +\dfrac{b (q_1 ^2 + 6 q_1  q_2  + q_2 ^2)}{q_1^2 q_2 ^2}
+\dfrac{c q_1 ^2 q_2 ^2 }{(q_1  + q_2 )^2}\,,
\\ \\
U_1=\dfrac{40a (q_1  - q_2 )^2 (q_1 ^2 + 10 q_1  q_2  + q_2 ^2)}{q_1 ^4 q_2 ^4} + \dfrac{5b (q_1  - q_2 )^2}{q_1 ^2 q_2 ^2}\,,
\\ \\
U_2=\dfrac{20a (q_1 ^2 + 6 q_1  q_2  + q_2 ^2) (q_2  + q_1 )^2}{q_1 ^3 q_2 ^3(q_1  - q_2 ) } - \dfrac{5 c q_1 ^3 q_2 ^3}{ (q_1  - q_2 ) (q_2  + q_1 )^2}\,,
\end{array}
\]
and
\[
\begin{array}{l}
\hat{V}=\dfrac{a (q_1 ^4 + 6 q_1 ^3 q_2  + 6 q_1 ^2 q_2 ^2 + 6 q_1  q_2 ^3 + q_2 ^4)}{q_1 ^4 q_2 ^4}+\dfrac{b(3 q_1 ^2 + 14 q_1  q_2  + 3 q_2 ^2)}{q_1  q_2  \sqrt{q_1 ^2 + 6 q_1  q_2  + q_2 ^2}}
+\dfrac{c (q_1 ^2 + 6 q_1  q_2  + q_2 ^2)}{q_1 ^2 q_2 ^2}\,,
\\ \\
\hat{U}_1= -\dfrac{5a (q_1  - q_2 )^2  (q_1 ^2 + 6 q_1  q_2  + q_2 ^2) (q_1  + q_2 )^2}{q_1 ^4 q_2 ^4}-\dfrac{20b (q_1  - q_2 )^2 \sqrt{q_1 ^2 + 6 q_1  q_2  + q_2 ^2}}{q_1  q_2 }\,,
\\ \\
\hat{U}_2=\dfrac{ 5a (q_1  + q_2 )}{q_1 ^2 q_2 ^2 (q_1  - q_2 )}-\dfrac{5b (q_1  + q_2 )}{2\sqrt{q_1 ^2 + 6 q_1  q_2  + q_2 ^2} (q_1  - q_2 )}  - \dfrac{5 c (q_1  + q_2 )}{2 q_1  q_2  (q_1  - q_2 )}\,.
\end{array}
\]
In $x,y$-variables potentials are equal to
\ben\label{V18}
V&=&\dfrac{16a(16x^4 - 28x^2y^2 + 11y^4)}{x^8} -\dfrac{4b(x^2 - y^2)}{x^4}  +\dfrac{c x^4}{4y^2}\,,\\
\nn\\
\label{V19}
\hat{V}&=&-\frac{4a(x^4 + 2x^2y^2 - 4y^4)}{x^8} + \frac{2b(2x^2 - 3y^2)}{x^2\sqrt{y^2-x^2}} -\frac{4c(x^2 - y^2)}{x^4}\,.
\en

We can  prove that  Hamiltonians $H_1$ (\ref{H})  and $\hat{H}_1$ (\ref{hH})  do not commute with polynomial of fourth order in momenta
\[H_3=L+\sum_{k=0}^4 \sum_{i=0}^k f_{ik}(q_1,q_2)p_1^{k-i}p_2^i\]
for any $f_{ik}(q_1,q_2)$. As above it means that the corresponding Hamiltonian vector fields admit quartic invariants only for geodesic motion.

\subsection{Case $m=4$ and $k={1}/{2}$}
In this case  geodesic Hamiltonian
\[
T=\frac{q_1^4\left(\frac{q_1}{2} + q_2\right)p_1^2}{q_1 -q_2} +\frac{q_2^4\left(q_1 +\frac{q_2}{2}\right)p_2^2}{q_2-q_1}
\]
commute with cubic polynomial
\[K=K_1K_2=\hat{K}_1\hat{K}_2=\frac{q_1^2q_2^2(q_1^2p_1 - q_2^2p_2)^2\Bigl(q_1^2(2q_1 + 3q_2)p_1 - q_2^2(3q_1 + 2q_2)p_2\Bigr)}{(q_1 -q_2)^3}\,,\]
where
\[
K_1= q_1 ^2 (2 q_1  + 3 q_2 ) p_1 -q_2 ^2  (3 q_1  + 2 q_2 ) p_2\,,\qquad\mbox{and}\qquad
\hat{K}_1=(q_1 ^2p_1  -  q_2 ^2p_2)\,.
\]
As a result, we obtain  two integrable systems  (\ref{H}-\ref{hH}) with
\[
\begin{array}{rcl}
V&=&\dfrac{a(q_1 ^2 + 3 q_1  q_2  + q_2 ^2) (3 q_1 ^2 + 8 q_1  q_2  + 3 q_2 ^2)}{q_1 ^4 q_2 ^4} + \dfrac{b (q_1 ^2 + 3 q_1  q_2  + q_2 ^2)}{q_1 ^2 q_2 ^2}+\dfrac{c q_1 ^2 q_2 ^2}{(q_1  + q_2 )^2}\,,
\\ \\
U_1&=&-\dfrac{ 2a (q_1  - q_2 )^2 (q_1 ^2 + 3 q_1  q_2  + q_2 ^2)^2}{q_1 ^4 q_2 ^4}-\dfrac{2c q_1 ^2 q_2 ^2 (q_1  - q_2 )^2}{(q_1  + q_2 )^2} \,,
\\ \\
U_2&=&\dfrac{2a (q_1 ^2 + 4 q_1  q_2  + q_2 ^2)}{q_1 ^2 q_2 ^2 (q_1  - q_2 )} +\dfrac{2 b}{q_1  - q_2 }\,,
\end{array}
\]
and
\[
\begin{array}{l}
\hat{V}=\dfrac{a(17 q_1 ^4 + 108 q_1 ^3 q_2  + 198 q_1 ^2 q_2 ^2 + 108 q_1  q_2 ^3 + 17 q_2 ^4)}{q_1 ^4 q_2 ^4}
+\dfrac{b (7 q_1 ^2 + 18 q_1  q_2  + 7 q_2 ^2) }{q_1 ^2 q_2 ^2} +\dfrac{c(q_1+q_2)}{q_1q_2}\,,
\\ \\
\hat{U}_1=\dfrac{16a (q_1  - q_2 )^2 (q_1 ^2 + 4 q_1  q_2  + q_2 ^2)}{q_1 ^4 q_2 ^4} + \dfrac{8b (q_1  - q_2 )^2}{q_1 ^2 q_2 ^2}\,,
\\ \\
\hat{U}_2=\dfrac{8a (q_1  + 3 q_2 ) (3 q_1  + q_2 ) (q_1  + q_2 )}{q_1 ^2 q_2 ^2 (q_1  - q_2 )} + \dfrac{4b (q_1  + q_2 )}{q_1  - q_2}  + \dfrac{2 c q_2  q_1}{q_1  - q_2 }\,.
\end{array}
\]
In $x,y$-variables potentials are equal to
\ben
\label{V20}
V&=&\frac{2a(x^2 -4y^2)(x^2 - 6y^2)}{x^8} -\frac{b (x^2 - 4y^2)}{x^4} - \frac{cx^4}{4y^2}\,,\\
\nn\\
\label{V21}
\hat{V}&=&\frac{16a(x^4 - 10x^2y^2 + 17y^4)}{x^8} -\frac{4b(x^2 - 7y^2)}{x^4} -\frac{2cy}{x^2}\,.
\en
Thus, we present 21 integrable systems with sextic invariants associated with metric (\ref{gen-ham}) depending on coordinates.

Other nontrivial families of integrable systems with quartic invariants can be obtained similarly. Indeed, we can consider integrals of motion
\bq\label{gen-L}
H_1=T+V(q_1,q_2)\,,\qquad H_2=\Bigl(K_1^2+{U}_1(q_1,q_2)\Bigr)\Bigr(L_2+{U}_2(q_1,q_2)\Bigr)
\eq
where ${V}$ and ${U}_{1,2}$ are solutions of the equation (\ref{m-eq}). For instance,  at $m=1$ and $k=2$
if we substitute Hamiltonian
\[
H_1=\frac{ q_1  (2 q_1  + q_2 ) p_1^2}{q_1  - q_2 } +\frac{ q_2  (q_1  + 2 q_2 ) p_2^2}{u_2-u_1} + V(q_1 , q_2 )
\]
and
\[
H_2=\Bigl((q_1^2p_1 -q_2^2p_2)^2 + U_1(q_1 , q_2 )\Bigr)\left(\frac{(q_1p_1^2 -q_2 p_2^2)}{(q_1  - q_2 )^3} + U_2(q_1 , q_2 )\right)
\]
into $\{H_1,H_2\}=0$ and solve the resulting equation, then we obtain the following solution
\[
V=a(q_1  + q_2 )\bigl(2 q_1 ^2 + q_1  q_2  + 2 q_2 ^2\bigr)  + b\bigl(2 q_1 ^2 + 3 q_1  q_2  + 2 q_2 ^2\bigr) +c (q_1  + q_2 )+\frac{d(q_1  + q_2 )}{q_1  q_2}+\frac{e (q_1  + q_2 )^3}{q_1 ^3 q_2 ^3}
\]
and integrable Hamiltonian
\[
H_1=\frac{y(p_x^2 + 2p_y^2)}{2}+\frac{xp_xp_y}{2} +2ay(3x^2 + 8y^2) +b  (x^2 + 8y^2) + 2cy-\frac{2d y}{x^2} + \frac{8ey^3}{x^6}
\]
depending on five parameters $a,b,c,d,$ and $e$. The corresponding Hamiltonian vector fields has the quartic invariant $H_2$.

\section{Conclusion}
In 1910 Darboux studied two-dimensional systems in the Euclidean plane with quadratic invariants \cite{darb2} and his calculations became one of the foundations of the modern theory of separation of variables in orthogonal curvilinear coordinates.

In 1935 Drach studied two-dimensional systems in the pseudo-Euclidean plane with cubic invariants \cite{dr35}. Seven of the ten
Drach potentials are superintegrable with two quadratic additional integrals \cite{ts00} and, therefore,  cubic Drach invariants exist due to
 the Abel and  Riemann-Roch theorems \cite{ts19,ts20}.

In \cite{ts11} we studied three families of  superintegrable two-dimensional systems in the pseudo-Euclidean planes, for which one additional first
integral is quadratic, and the second one can be arbitrarily polynomials of a high degree in momenta. Some of these systems were obtained using degenerate deformations of the Poisson brackets but we do not know of a mathematical justification for the existence of these systems till now.

In this note, we start with two-dimensional systems with a position-dependent mass having cubic invariants. Some of these systems are also superintegrable. In contrast with the Drach systems, these cubic invariants can be used to construct systems with sextic invariants. From the technical point of view, it is a consequence of  dependence of metrics on the coordinates.

In \cite{ts17} we obtain a triad of bivectors $\Pi_\pm$ and $\pi$, which form a Poisson bivector
\[
\Pi'=\frac{\Pi_-}{a}+\pi+a\Pi_+
\]
compatible with canonical Poisson bivector $\Pi$, so that
\[X=\Pi dH_1=\Pi' dH_2\]
for a system with sextic invariant at $m=4$ and $k=3$.  In the future, we plan to study similar triads of  bivectors $\Pi_\pm$ and $\pi$ and the corresponding bi-Hamiltonian vector fields $X$ for other integrable cases obtained in this note. We hope that this mathematical experiment allows us to get new results for natural Hamiltonian systems with other metrics.

The work was supported by the Russian Science Foundation (project 21-11-00141).

\end{document}